\documentclass[12pt]{iopart}
\usepackage{graphicx}
\usepackage{iopams}
 \usepackage{cite}
\begin{document}

\title{Bunches of Random Cross-correlated Sequences}

\author{A~A~Maystrenko$^1$, S~S~Melnik$^2$, G~M~Pritula$^2$ and O~V~Usatenko$^2$ }

\address{$^{1}$ V~N~Karazin Kharkiv National University,
4 Freedom Square, 61077, Kharkiv, Ukraine}

\address{$^{2}$ A Ya Usikov Institute for Radiophysics and
Electronics, Ukrainian Academy of Science, 12 Proskura Street, 61085
Kharkov, Ukraine}
\ead{pritula.galina@gmail.com}
\begin{abstract}

Statistical properties of random cross-correlated sequences constructed by 
the convolution method (likewise referred to as the Rice's or the inverse 
Fourier transformation) are examined. Algorithms for their generation are 
discussed. They are frequently reduced to solving the problem for 
decomposition of the Fourier transform of the correlation matrix into a 
product of two mutually conjugate matrices; different decompositions of 
the correlation matrix are considered.  The limits of weak and strong 
correlations for the one-point probability and pair correlation functions 
of the sequences are studied. Special cases of heavy-tailed distributions 
resulting from the convolution generation are analyzed. Anisotropic 
properties of statistically homogeneous random sequences related to 
asymmetry of a filtering function are discussed.

\end{abstract}
\submitto{\JPA}
\maketitle

%%%%%%%%%%%%%%%%%%%%%%%%%%%%%%%%%%%%%%%%%%%%%%%%%%%%%%%%%%%%%%%%%%
\section{Introduction}\label{Intro}
%%%%%%%%%%%%%%%%%%%%%%%%%%%%%%%%%%%%%%%%%%%%%%%%%%%%%%%%%%%%%%%%%%
Over the past several decades the correlated disorder has been the
focus of a large number of studies in different fields of science.
The unflagging interest in systems with correlated fluctuations
is explained by specific properties they demonstrate and their
prospective applications. Moreover, at present there is a commonly
accepted viewpoint that our world is complex and correlated. The
most peculiar manifestations of this concept are the records of
brain activity and heart beats, human and animal communication,
written texts, DNA and protein sequences, data flows in computer
networks, stock indexes, etc.

The studies of random systems in physical and engineering sciences
can be divided into two parts. The first one investigates, analyzes
and predicts the behavior of such systems, whereas the second one,
which is considerably smaller, develops the methods of construction,
or generation, of random processes with desired statistical
properties. The essence of the second approach is to construct a
mathematical object (for example, a correlated  sequence of symbols
or numbers) with tailored statistical characteristics.
This approach provides not only  a deeper insight into the nature of
correlations but also a creative tool for designing the devices and
appliances with random components in their structure such as
different wave-filters, diffraction gratings, artificial
materials, antennas, converters, delay lines, etc. These devices
can exhibit unusual properties or anomalous dynamical, kinetic or
transport characteristics controlled by a proper choice of disorder.

There are many algorithms for generating long-range correlated
sequences: the Mandelbrot fast fractional Gaussian noise
generation~\cite{Mand71}, the Voss procedure of consequent random
addition~\cite{voss}, the correlated L\'evy walks~\cite{shl}, the
expansion-modification Li method~\cite{Li89}, the method of Markov
chains~\cite{RewUAMM}, etc. We believe that the convolution method
(and its variant --- the Fourier filtering method~\cite{czir}) is
one of the most efficient. This
method may be used to generate enhanced diffusion, isotropic and
anisotropic self-affine surfaces, isotropic and anisotropic
correlated percolation~\cite{Makse}. The convolution
method allows one to construct sequences with random elements belonging
to a continuous space of states ---  the space of real numbers
$\mathbb{R}=(-\infty,\infty)$ --- the widest possible space. Note that if some
restrictions on possible states of random variables are imposed,
say, we need to generate a random dichotomous sequence, then the problem
becomes much more
complicated~\cite{CarpStan1,CarpStan2,hod,nar1,nar2,UYa,IKMU,gener}.

In the present paper we generalize the convolution method of generating a
discrete statistically homogeneous colored sequence with a given
correlation function. The method is based on a linear transformation
of white noise with the use of the filtering function -- the
kernel of the convolution operator -- and gives a rather simple
relation between this function and the pair correlation function
\cite{IzrKrMak}. Here we present the matrix generalization of this
method to construct a bunch of $N$ cross--correlated sequences and
study their statistical properties.

The scope of the paper is as follows. First, we discuss briefly the
Rice convolution method for generating random sequences.   In
section~\ref{GenConv} we generalize the method  to a set (or, a
bunch) of $N$ cross-correlated statistically homogeneous sequences
with a prescribed binary correlation matrix. Some analytical
solutions for the problem of the correlation matrix decomposition
are presented in section \ref{ExplSol}. Section \ref{CharFunc}  is
devoted to studying statistical properties of sequences constructed
with this particular method. Section~\ref{Ex-Gen} contains an
example for constructing two cross-correlated chains with a given
correlation matrix.
%
%%%%%%%%%%%%%%%%%%%%%%%%%%%%%%%%%%%%%%%%%%%%%%%%%%%%%%%%%%%%%%%%%%%%%%%%%%%%%%%
\section{Introduction to Convolution Method}\label{ConvIntro}
%%%%%%%%%%%%%%%%%%%%%%%%%%%%%%%%%%%%%%%%%%%%%%%%%%%%%%%%%%%%%%%%%%%%%%%%%%%%%%

This section provides a brief introduction to the most known
and frequently used method for generating random correlated
sequences with a continuous space of states~\cite{Rice4454,Rice,Saupe88Feder88,
WOD95,IzKr99,IzMak05,Grigo06, Garcia,Czir95,Makse,IKMU,IzrKrMak,RewUAMM}.

Let us introduce a homogenous random white noise sequence
$\{\xi(n)\}$ of independent and identically distributed (i.i.d.)
variables $\xi(n) \in \mathbb{R}$, $n \in
\mathbb{Z} = (..., -2,-1,0,1,2, ...)$. All statistical properties of
the sequence are determined by the one-point probability
distribution function (PDF) and its moments. The most important
among them are the mean value $\langle\xi(n)\rangle$, which we put hereafter
equal to zero without loss of generality, and the two-point
correlation function, which is expressed via the unit variance
$\sigma_{\xi}^2$:
\begin{equation}\label{118}
\langle\xi(n)\rangle=0, \, \, C_{\xi}(r) = \langle\xi(n) \xi(n+r)
\rangle=\sigma_{\xi}^2 \,
\delta_{r,0},\, \,\sigma_{\xi}=1,
\end{equation}
where $\delta_{r,0}$ is the Kronecker delta symbol. The brackets
$\langle...\rangle$ mean a statistical (arithmetic, Ces\`{a}ro's) average along
the chain,
\begin{equation}\label{mean arithm}
\langle f(\xi(n))\rangle= \lim_{M\rightarrow \infty}\frac{1}{2M+1}
\sum_{n=-M}^{M}f(\xi(n)),
\end{equation}
or the equivalent average with the PDF  $\rho_{\Xi}(\xi)$
\begin{equation}\label{mean distr}
\langle f(\xi(n))\rangle=
 \int \! \! d\xi \, \rho_{\Xi}(\xi)\,f(\xi).
\end{equation}
It is supposed that the mean values and the variance of sequence
$\{\xi(n)\}$ exist.

The linear convolution transformation with \emph{filtering} function
$F(n)$ generates a new  \emph{correlated} sequence $\{x(n)\}$,
\begin{equation}\label{x F alpha}
x(n) = \sum_{n'=-\infty}^{\infty}  F(n-n')\, \xi (n') .
\end{equation}
This formula (probably, the most important throughout the paper, no
matter how simple it may seem) determines both analytical properties
of correlated sequence and the method of its numerical construction
(beginning with the white-noise sequence $\{\xi(n)\}$). So, we have to be
able to answer a number of questions: what restrictions should be
imposed on the filtering function, what  one-point probability
distribution and two-point, or pair, correlation functions of
$\{x(n)\}$ sequence are.

It is evident from the first formula of
equation~(\ref{118}) that
\begin{equation}\label{mean x}
\langle x(n)\rangle= 0.
\end{equation}
It is also simple to calculate the pair auto-correlation function%
\begin{equation}\label{C sum F}
C_x(r) = \sum_{n=-\infty}^{\infty} F(n+r)F(n).
\end{equation}
This equation is readily derived by substituting of equation~(\ref{x F
alpha}) into the definition  of correlation function $C_x(r)$,
\begin{eqnarray}
 C_{x}(r>0) &=&
 \left\langle
  \left( x(n+r)-\langle x \rangle \right)\left( x(n)-\langle x \rangle \right)
\right\rangle
\nonumber
\\ &=&
\lim_{M\rightarrow \infty}\frac{1}{2M-r+1}
\sum_{n=-M}^{M-r} x(n+r)x(n)=C_{x}(-r).
\label{def_autocorr}
\end{eqnarray}
Since our main purpose is to consider the cross-correlated
sequences, it is worth to note that sequences $\{\xi(n)\}$
and $\{x(n)\}$ are correlated,
\begin{equation}\label{def_corr}
 C_{\xi x}(r)=
 \left\langle \xi(n) x(n+r)\right\rangle= F(r).
\end{equation}
This property explains the meaning of the filtering function.

Considering the sets of functions $F(n)$ and $F(n+r)$ as two
vectors and their combination $\sum_{n=-\infty}^{\infty} F(n+r)F(n)$
as a scalar product of two equal vectors, one of which rotates
around another (from the passive point of view the components $F(n+r)$ are
obtained by cyclic rotations of the coordinate system and $F(n)$ are
the components of the vector prior to rotation) we conclude that
\begin{equation}\label{IneqRot}
C_x(0) \geq C_x(r).
\end{equation}
We will also use the \emph{correlation coefficient} $K_x(r)$,
\begin{eqnarray}\label{CM-Cor-beta}
K_x(r)= \frac{C_x(r)}{C_x(0)}.
\end{eqnarray}
By definition, the correlation coefficient $K_x(r)$ is normalized to
unity, $K_x(0)=1$. The last property can be seen as a transformation
$x(n)\rightarrow x(n)/ \sqrt{C_x(0)}$, which renormalizes the old variables
$x(n)$ to the new ones with unit variances. Because the
initial uncorrelated chain $\{\xi(n)\}$ is statistically homogeneous
and the generating function $F(.)$ in equation~(\ref{x F alpha}) depends on
the difference $n-n'$ only, the generated random sequence $\{x(n)\}$
is statistically homogeneous as well. This property implies the
independence of one-point distribution functions on the number of
site, the possibility of averaging~(\ref{mean distr})
along the sequence, the dependence of binary correlation functions on
the difference of their arguments and many other useful properties
of the sequence.

Thus, equation~(\ref{C sum F}) relates the pair correlation function to
the filtering function provided that the series
$\sum_{n=-\infty}^{\infty} F(n+r)F(n)$ converges. A simpler
relation between them, the Fourier transformation of equation~(\ref{C sum
F}), reads
\begin{equation}\label{K_F_1chain}
\tilde{C}_x(k) = \tilde{F}(k) \tilde{F}(-k).
\end{equation}
Here we use the following formulae for the Fourier transform and its inverse:
\begin{equation}\label{FourTr1}
 \tilde{G}(k)=
  \sum_{r=-\infty}^{\infty}\!\!
  G(r)\exp (-ikr),
  \qquad
  G(r)=
  \frac{1}{2\pi}
  \int\limits_{-\pi}^{\pi} \! \! dk \, \tilde{G}(k) \exp(ikr).
 \end{equation}
Two properties of the function $ \tilde{G}(k)$, stemming from
discreteness and real-valuedness of  the function $G(r)$, will be useful in
what follows:
\begin{equation}\label{Df-Ds}
 \tilde{G}(k+2\pi)= \tilde{G}(k), \qquad  \tilde{G}(-k)= \tilde{G}^\ast(k).
\end{equation}
From the second expression of equation(\ref{Df-Ds}) and equation~(\ref{K_F_1chain}),
 we immediately obtain
the Wiener-Khinchin theorem~\cite{Brockwell} for power spectrum
$\tilde{C}_x(k)$,
\begin{equation}\label{WinKhin}
\tilde{C}_x(k) = \tilde{F}(k) \tilde{F}(-k)=|\tilde{F}(k)|^2 \geq 0.
\end{equation}
It is easy to see that equations~(\ref {C sum F}) and (\ref {K_F_1chain})
correctly reflect the parity of function $ C_x (r) $ and its
Fourier transform $ \tilde {C}_x(k) $ for any function $ F (n) $:
\begin{equation}\label{Pair prop}
C_x(-r)=C_x(r),\,\,\,\, \tilde{C}_x(-k)=\tilde{C}_x(k) .
\end{equation}
The solution of equation~(\ref{K_F_1chain}) is %
\begin{equation}\label{F k C}
 \tilde{F}(k) =   \exp [i\varphi(k)]\,\sqrt{\tilde{C}_x(k)},
\end{equation}
where $\varphi(k)$ is an arbitrary odd function,
$\varphi(-k)=-\varphi(k)$.

Thus, the solution of the problem of constructing a random sequence
with a given correlation function $ C_x(r) $, or its Fourier
transform $ \tilde {C}_x(k) $, is reduced to finding the filtering
function $ F(n) $,  which determines (see equation~(\ref{x F alpha})) the
transformation of uncorrelated sequence $ \{ \xi (n) \}$ into
correlated $ \{x(n) \} $-sequence.

For numerical generation of random sequences, in equation~(\ref{x F alpha}) an
even kernel function $ F (n)$ is commonly used.
Nevertheless, we see that equation~(\ref{F k C}) allows one to find
solutions in more general form. Let us consider this in more detail and
represent the filtering function $ F (n) $ as the sum of its even
$ F_c (n) $ and odd $ F_s (n) $ parts, $ F = F_c (n) + F_s (n) $. Then
equations~(\ref {C sum F}) and (\ref {K_F_1chain}) become
\begin{eqnarray}
C_x(r) = \sum_{n=-\infty}^{\infty} [F_c(n+r)F_c(n)+F_s(n+r)F_s(n)],
\nonumber
\\
\quad \tilde{C}_x(k) = \tilde{F_c}^2(k) + \tilde{F_s}^2(k).
\label{K_F_1cs}
\end{eqnarray}
Here $ \tilde {F_c} (k) $ and $ \tilde {F_s} (k) $ are the Fourier
cosine and sine transforms of $ F_c (n) $ and $ F_s (n) $,
respectively.

Another method, the inverse Fourier transformation, for generating a
sequence of random numbers with long-range correlations is given in
\cite{Makse}. This method can be viewed as a modification of the
above discussed convolution method and it is based on the Fourier
transform of equation~(\ref{x F alpha})
\begin{equation}\label{x F alpha2}
 \tilde{x}(k) =  \tilde{F}(k) \tilde{\xi}(k).
\end{equation}
The first step in generating correlated random numbers is to
calculate the Fourier transform of uncorrelated sequence $ \{ \xi
(n) \}$. A method which enables one to avoid these cumbersome
calculations and generate directly the values of $ \tilde{\xi}(k)$
is presented in appendix.

Now consider the effect of the filtering function shape on
correlation properties of a random sequence qualitatively. Suppose
that the filtering function is bell-shaped
with a characteristic scale of the order of unity. The
characteristic scale of the function $F(n/R_c))$ is then $R_c>0$.
For $r=0$ the overlap of functions $F(n)$ and $F(n+r)$ in
equation~(\ref{C sum F}) is maximal, so that $C_x(r=0)$ is maximal  as
well. If the ``distance'' $r$ between $F(n)$ and $F(n+r)$ exceeds
$R_c$, the overlap almost vanishes, so that $C_x(r>R_c)$ takes on
small values. It means that, by an order of magnitude, the
characteristic scale of the function $F(n/R_c))$ is, at the same time,
the correlation length $R_c$ of the generated random sequence.
Furthermore, it is clear that if $R_c$ goes to zero, the sequence
$\{x(n)\}$ becomes uncorrelated white noise with
$K_x(r)=\delta_{r,0}$ and $\widetilde{K}_x(k)=1$. The other limit,
$R_c$ goes to infinity, describes totally correlated sequence,
$K_x(r)=1$ and $\widetilde{K}_x(k)=2\pi \delta(k)$. All the
above-mentioned facts are demonstrated by the following simple
example~\cite{Antenna}:
\begin{equation}\label{Exam Kr}
K(r))=\frac{1}{1-\exp(-\pi R_c)}\frac{1-(-1)^r \exp(-\pi R_c)
}{1+(r/R_c)^2},
\end{equation}
\begin{equation}\label{Exam Kk}
 \tilde{K}(k))=\frac{\pi R_c}{1-\exp(-\pi R_c)} \exp(- R_c |k|),\quad \tilde{F}(k))=
 \sqrt{\tilde{K}(k))}.
\end{equation}

Note, if (in some cases) at  $r>R_c$ the filtering function
vanishes, then the correlation function also vanishes at $r>2R_c$,
$C_x(r>2R_c)=0$.

%%%%%%%%%%%%%%%%%%%%%%%%%%%%%%%%%%%%%%%%%%%%%%%%%%%%%%%%%%%%%%%%%%%%%%%%%%%%%%%
\section{Generalization of Convolution Method}\label{GenConv}
%%%%%%%%%%%%%%%%%%%%%%%%%%%%%%%%%%%%%%%%%%%%%%%%%%%%%%%%%%%%%%%%%%%%%%%%%%%%%%

The convolution method outlined in the previous section can be generalized to
the generation of
a set of $N$ cross-correlated statistically homogeneous random sequences
$\{x_{i}(n)\}$,
$x_i(n) \in\mathbb{R}$, $n \in \mathbb{Z}$, with a given binary
correlation matrix $\mathbf C(r)$,  whose entries $C_{ij}(r)$,
$i,j=1,2,...,N$, are
\begin{equation}
  C_{ij}(r)=
  %\left
    \langle
    \left( x_{i}(n+r)-\langle x_{i} \rangle \right)\left( x_{j}(n)- \langle x_{j}
    \rangle \right)
 %\right
\rangle.
  \label{def_ncorr}
 \end{equation}
The diagonal elements of the correlation matrix are the auto-correlation
functions, which describe the relationships between the elements of
the same sequence, while the non-diagonal entries represent
cross-correlations between the elements of different sequences. As
above, here we also suppose $\langle x_i \rangle=0$,
$i={1,...,N}$.

The correlation matrix elements are real and, as
seen directly from equation~(\ref{def_ncorr}),  have the following
property:
\begin{equation}
C_{ij}(r)=C_{ji}(-r). \label{corr_element}
\end{equation}
In terms of the Fourier transform determined by equation~(\ref{FourTr1}),
this property reads
\begin{equation}
\tilde{C}_{ij}(k)=\tilde{C}_{ji}(-k)={\tilde{C}_{ji}^{*}(k)},
\label{ft_corr_element}
\end{equation}
where the asterisk denotes complex conjugation. Equations~
(\ref{corr_element}) and (\ref{ft_corr_element}) can be written in
matrix form:
\begin{equation}
\mathbf{C}(r)=\mathbf{C}^{\top}\!(-r), \qquad
\tilde{\mathbf{C}}(k)=\tilde{\mathbf{C}}^{\top}\!(-k)=
\tilde{\mathbf{C}}^{\dagger}\!(k), \label{corr_matrix}
\end{equation}
where the symbols ${^\top}$ and $^{\dagger}$ indicate the transpose
and conjugate transpose of a matrix, respectively. Since the matrix
$\tilde{\mathbf{C}}(k)$ has complex entries and is equal to its
conjugate transpose, it is Hermitian; hence, its diagonal
elements are real.

To construct the correlated sequences $\{x_{i}(n)\}$, let us consider as a
starting point $N$ independent uncorrelated white-noise random
sequences $\{\xi_i\ (n)\}$:
\begin{equation}
 \langle \xi_{i} \rangle=0,
   \qquad
 \langle \xi_{i} (n) \xi_{j} (n') \rangle = \delta_{ij} \delta_{nn'}.
    \label{def_alpha_i}
 \end{equation}

Similarly to the 1-sequence convolution method, we construct $N$
correlated sequences $\{x_{i}(n)\}$ as a sum of convolutions of
delta-correlated sequences $\{\xi_{j} (n)\}$ with  filtering
functions $F_{ij}(n)$ in the following way:
\begin{equation}
  x_{i}(n)=
    \sum_{j=1}^{N}\sum_{n'=-\infty}^{\infty}\!\!
  F_{ij}(n-n') \xi_{j}(n').
    \label{def_x_i}
   \end{equation}

Substituting equation (\ref{def_x_i}) into definition of correlator
(\ref{def_ncorr}) and using property (\ref{def_alpha_i}), we reveal
the relationship between the elements of the correlation matrix and
the filtering functions:
    \begin{eqnarray}
    C_{ij} (r) & = &
        \sum_{p=1}^{N}\sum_{n=-\infty}^{\infty}\!\!
    F_{ip}(n+r)F_{jp}(n).
    \label{ncorr_r}
    \end{eqnarray}

 The Fourier transform translates equation(\ref{ncorr_r}) into the system of equations
 in $k$-space:
    \begin{eqnarray}
    \tilde{C}_{ij}(k) & = &
            \sum_{p=1}^{N}\tilde{F}_{ip}(k)\tilde{F}_{jp}(-k).
    \label{ncorr_k}
    \end{eqnarray}

The matrix form of equation~(\ref{ncorr_k}), which generalizes
equation (\ref{K_F_1chain}) to the case of $N$ cross-correlated sequences,
embodies the algebraic content of the problem under consideration:
\begin{equation}
    \tilde{\mathbf{C}}(k)=\tilde{\mathbf{F}}(k)\tilde{\mathbf{F}}^{\top}(-k)
    \label{eq_matrix}
\end{equation}
or, equivalently,
\begin{equation}
    \tilde{\mathbf{C}}(k)=\tilde{\mathbf{F}}(k)\tilde{\mathbf{F}}^{\dagger}(k).
    \label{eq_matrixHermConjug}
\end{equation}

Thus, to construct the  bunch of $N$ cross-correlated sequences
$\{x_{i}(n)\}$ with the given correlators we have to find the
factorization of the Hermitian matrix $\tilde{\mathbf{C}}(k)$ into a
product of the Fourier transform of generating function
$\tilde{\mathbf{F}}(k)$ and its Hermitian transpose
$\tilde{\mathbf{F}}^{\dagger}(k)$. This is the well-known problem of
liner algebra (see, for example, \cite{HornJohnson}) and there are
different approaches to its solution. Let us consider some of them
in relation to the problem under consideration.

\textbf{Spectral decomposition}. Since the correlation matrix
$\tilde{\mathbf{C}}(k)$ is Hermitian, it can be diagonalized by a
unitary matrix $\mathbf{U}$ and the resulting diagonal matrix has
real entries only (\cite{HornJohnson}, Theorem 4.1.5). If the matrix
$\tilde{\mathbf{C}}$ is positive-definite, we can easily find the
formal solution of equation(\ref{eq_matrixHermConjug}):
\begin{equation}
    \tilde{\mathbf{C}}= \mathbf{U}\Lambda \mathbf{U}^{\dagger},
        \qquad
        \Lambda=\mbox{diag}\left(\lambda_1,...,\lambda_N \right).
        \label{matrixU}
 \end{equation}
Here $\lambda_i\geq 0$ are the eigenvalues of matrix
$\tilde{\mathbf{C}}(k)$. This implies
 \begin{equation}
    \tilde{\mathbf{F}}(k)= \mathbf{U}\sqrt{\Lambda},
        \qquad
        \sqrt{\Lambda}=\mbox{diag}\left(\sqrt{\lambda_1},...
        \sqrt{\lambda_N}\right).
    \label{spectral}
\end{equation}

\textbf{Cholesky decomposition}. For Hermitian positive-definite
matrices, there are other decompositions, which solve
equation (\ref{eq_matrixHermConjug}). One of them is the Cholesky
decomposition factorizing the matrix into a lower triangular matrix
$\mathbf{L}$ with strictly positive diagonal entries and its
conjugate transpose \cite{GilbertStew, Watkins, NumRecipes},
\begin{equation}
            \tilde{\mathbf{C}}= \mathbf{L}\mathbf{L}^{\dagger},
        \label{chol}
\end{equation}
which immediately provides the solution to our problem,
$\tilde{\mathbf{F}}=\mathbf{L}$.

\textbf{LDL factorization}. Besides, one can use the so-called LDL
decomposition factorizing a Hermitian matrix into a lower triangular
matrix $\mathbf{L}$, a diagonal matrix $\mathbf{D}$ with positive
entries and conjugate transpose of the lower triangular matrix
\cite{NumRecipes}),
\begin{equation}
            \tilde{\mathbf{C}}= \mathbf{L}\mathbf{D}\mathbf{L}^{\dagger}.
        \label{ldl}
\end{equation}
In the context of our problem,
$\tilde{\mathbf{F}}=\mathbf{L}\sqrt{\mathbf{D}}$.

\textbf{Hermitian ansatz.} It is also natural to look for a solution
of problem (\ref{eq_matrixHermConjug})
assuming $\tilde{\mathbf{F}}(k)$ to be a Hermitian matrix,
$\tilde{\mathbf{F}}(k)=\tilde{\mathbf{F}}^{\dagger}\!(k)$. In this
case equation (\ref{eq_matrixHermConjug}) can be converted to
\begin{equation}
\tilde{\mathbf{C}}(k)=\tilde{\mathbf{F}}^{2}(k).
\label{eq_matrix_partic}
\end{equation}
Formally the solution of this equation can be presented as
\begin{equation}
\tilde{\mathbf{F}}=\mathbf{U}\sqrt{\Lambda}\:\mathbf{U}^{\dagger}
\label{formalsolution}
\end{equation}
with unitary matrix $\mathbf{U}$ and $\Lambda$ determined in
equations~(\ref{matrixU}) and (\ref{spectral}).

Note that all of the above discussed solutions are particular ones. The
general solution can be obtained from any of them by right
multiplication by an arbitrary unitary matrix $\mathbf{W}$; if
$\tilde{\mathbf{F}}$ is a solution of our problem, then such is
$\tilde{\mathbf{F}}^{G}=\tilde{\mathbf{F}}\mathbf{W}$. Thus, for
example, implementing the Hermitian ansatz and representing the
unitary matrix $\mathbf{W}$ as an exponential function of an
arbitrary skew-Hermitian matrix $\mathbf{A}$,
\begin{equation}
\mathbf{W}=\exp{\mathbf{A}},\quad \mbox{where} \quad
{\mathbf{A^{\dagger}}}=-\mathbf{A}, \label{unitary_and_A}
\end{equation}
we can write the general solution of equation (\ref{eq_matrixHermConjug})
as
\begin{equation}
\tilde{\mathbf{F}}^{G}(k)={\tilde{\mathbf{F}}(k)}\exp{\mathbf{A}(k)},\quad
\quad {\mathbf{A^{\dagger}}(k)}=-\mathbf{A}(k). \label{general}
\end{equation}
This solution is a matrix generalization of equation(\ref{F k C}) for the
problem of $N$ cross-correlated sequences.

Considered algorithms of decompositions (\ref{spectral}) --
(\ref{formalsolution}) are widely used in programming \cite{Deift}
and continue to be developed and optimized for specific forms of
matrices. Nevertheless, explicit analytical solutions of this problem
can be found just in a few situations. In the next section we are
going to discuss some of them.

%%%%%%%%%%%%%%%%%%%%%%%%%%%%%%%%%%%%%%%%%%%%%%%%%%%%%%%%%%%%%%%%%%%%%%%%%%%%

\section{Explicit solutions}\label{ExplSol}

%%%%%%%%%%%%%%%%%%%%%%%%%%%%%%%%%%%%%%%%%%%%%%%%%%%%%%%%%%%%%%%%%%%%%%%%%%%%%
Equation (\ref{eq_matrixHermConjug}) admits explicit solutions in
the case of cyclic bunch of $N$ statistically identical sequences $\{x_{i}(n)\}$
with the nearest neighbor cross-correlations, when the correlation
matrix entries $\tilde{C}_{ij}(k)$ are
\begin{equation}\label{corrcycled}
  \tilde{C}_{ij}(k)=
    \left\{
    \begin{array}{ll}
        A_i, & i=j,\\
    B^*_i, & j=i+1 \pmod N,\\
    B_i, & i=j+1 \pmod N.
    \end{array}
        \right.
\end{equation}

We consider the simplest case of $A_i=A$, $B_i=B$ and, hence, the
correlation matrix is
\begin{equation}
\tilde{\mathbf{C}}(k)=
  \pmatrix{
        A      & B          & 0                 &   \cdots & 0     & B^*     \cr
        B^*    & A          & B                 &          & 0       & 0       \cr
        0      & B^*        & \ddots            &   \ddots &             & \vdots  \cr
        \vdots &            & \ddots            & \ddots &B          & 0  \cr
        0      & 0          &                   &   B^*      &A          & B       \cr
        B      & 0          &\cdots             &    0     &  B^*  & A
    }.
\end{equation}
Under this assumption, one can verify by direct substitution that
\begin{equation}
  \tilde{\mathbf{F}}(k)=
    \pmatrix{
        a      & B/a        & 0                 &   \cdots & 0        & 0     \cr
        0      & a          & B/a               &          & 0        & 0       \cr
        0      & 0          & \ddots            &   \ddots &             & \vdots  \cr
        \vdots &            & \ddots            &   \ddots & B/a          & 0  \cr
        0      & 0          &                   &   0      & a          & B/a       \cr
        B/a    & 0          &\cdots             &    0     & 0           & a
    }
  \label{solcycled}
\end{equation}
is one of the solutions of equation (\ref{eq_matrixHermConjug}). Here
 \begin{equation}\label{a}
a= \left\{
  \frac{A}{2} \pm
    \sqrt{ \frac{A^2}{4}-\left|B\right|^2 }
\right\}^{1/2}, \qquad A^2\geq 4|B|^2.
\end{equation}

Multiplying matrix $\tilde{\mathbf{F}}$ by an arbitrary unitary
matrix, we get the general solution.

Another instance when the filtering matrix $\tilde{\mathbf{F}}(k)$
can be found explicitly, is the generation of two correlated
sequences $\{x_{1}(n)\}$  and $\{x_{2}(n)\}$, i.e. $N=2$ . The
particular case of the $2\times 2$ problem was considered in \cite{HerIzrTes},
where a solution was obtained for the special form of filtering
matrix
\begin{equation}\label{Eta}
\mathbf{F}(r)=
\pmatrix{
    {G_1}(r)\cos\eta \quad & \quad {G_1}(r)\sin\eta \cr
    {G_2}(r)\sin\eta \quad & \quad {G_2}(r)\cos\eta
},
\end{equation}
$\eta$ is the real parameter and ${G_1}(r)$, ${G_2}(r)$ are even
filtering functions for the auto-correlation functions $C_{11}$ and $C_{22}$.
This form of $\mathbf{F}$ implies the specific form of
cross-correlation function
\begin{equation}
C_{12}(k) \propto \sqrt{C_{11}(k)C_{22}(k)}.
\end{equation}

Now we will discuss the problem for a general form of
$\tilde{\mathbf{F}}(k)$. In the $2\times 2$ case
equation~(\ref{eq_matrixHermConjug}) is reduced to the system of three
equations
\begin{equation}
  \left\{
  \begin{array}{lcl}
  \tilde{C}_{11} & = & |\tilde{F}_{11}|^2+|\tilde{F}_{12}|^2,\\
  \tilde{C}_{22} & = & |\tilde{F}_{21}|^2+|\tilde{F}_{22}|^2,\\
  \tilde{C}_{12} & = &
  \tilde{F}_{11}\tilde{F}^*_{21}+\tilde{F}_{12}\tilde{F}^*_{22}.
  \end{array}
  \right.
\label{sys2X2}
\end{equation}
Their general solution is
\begin{eqnarray}
\tilde{F}_{12} & = & \sqrt{\tilde{C}_{11}-|\tilde{F}_{11}|^2}\:
e^{i(\beta+\theta-\phi_1)},
\label{F12}\\
\tilde{F}_{21} & = & \sqrt{\tilde{C}_{22}-|\tilde{F}_{22}|^2}\:
e^{i(\alpha-\theta-\phi_2)}. \label{sol2X2}
\end{eqnarray}
Here
\begin{eqnarray}
  \cos{\phi_1} & = &
    \frac{|\tilde{C}_{12}|^2-\tilde{C}_{22}|\tilde{F}_{11}|^2+\tilde{C}_{11}
    |\tilde{F}_{22}|^2}{2|\tilde{C}_{12}||\tilde{F}_{11}|\sqrt{\tilde{C}_{22}
    -|\tilde{F}_{22}|^2}},
\label{phi1}\\
 \cos{\phi_2} & = &
\frac{|\tilde{C}_{12}|^2+\tilde{C}_{22}|\tilde{F}_{11}|^2-\tilde{C}_{11}|
\tilde{F}_{22}|^2}{2|\tilde{C}_{12}||\tilde{F}_{22}|\sqrt{\tilde{C}_{11}-|
\tilde{F}_{11}|^2}},
\label{phi2}
\end{eqnarray}
and $\theta$ is the argument of $\tilde{C}_{12}$. The functions
$\tilde{F}_{11}=|\tilde{F}_{11}|e^{i\alpha}$,
$\tilde{F}_{22}=|\tilde{F}_{22}|e^{i\beta}$ are arbitrary up to the
condition
\begin{equation}\label{condition}
  \left(\tilde{C}_{22}|\tilde{F}_{11}|^2+\tilde{C}_{11}|\tilde{F}_{22}|^2-|
  \tilde{C}_{12}|^2\right)\leq  4(\tilde{C}_{11}\tilde{C}_{22}-|
  \tilde{C}_{12}|^2)|\tilde{F}_{11}|^2|\tilde{F}_{22}|^2.
\end{equation}
Condition (\ref{condition}) stems from the restriction imposed on
the right-hand sides of (\ref{phi1}) and (\ref{phi2}): their modulus
should be less then unity.

Passing to the limit $\tilde{F}_{12}\rightarrow 0$, from general
$2 \times 2$ solution (\ref{F12}) -- (\ref{condition}) one
can derive the Cholesky decomposition (\ref{chol}) discussed in the
previous section:
\begin{equation}\label{Solution_Ch}
\tilde{F}_{11}  = \sqrt{\tilde{C}_{11}},\,\,\tilde{F}_{21}=
C_{21}/\sqrt{\tilde{C}_{11}},\,\,\tilde{F}_{22} =
\sqrt{\frac{\tilde{C}_{11}\tilde{C}_{22}-|\tilde{C}_{21}|^2}{\tilde{C}_{11}}},
\end{equation}
where $\tilde{C}_{11}\tilde{C}_{22}>|\tilde{C}_{12}|^2$.

The elegant explicit $2 \times 2$ solution can be found if the
filtering matrix $\tilde{\mathbf{F}}(k)$ is Hermitian,
$\mathbf{F}(k)=\mathbf{F}^{\dagger}\!(k)$, and equation (\ref{sys2X2}) is
converted to
\begin{equation}
  \left\{
  \begin{array}{lcl}
  \tilde{C}_{11} & = & \tilde{F}_{11}^2+|\tilde{F}_{12}|^2,\\
  \tilde{C}_{22} & = & |\tilde{F}_{12}|^2+\tilde{F}_{22}^2,\\
  \tilde{C}_{12} & = &
  \tilde{F}_{12}(\tilde{F}_{11}+\tilde{F}_{22}).
  \end{array}
  \right.
\label{sys2X2_hermit}
\end{equation}
In this particular case the solution is
\begin{equation}
\tilde{F}_{ij}=
\frac{\tilde{C}_{ij}+\delta_{ij}\sqrt{\tilde{C}_{11}\tilde{C}_{22}-
|\tilde{C}_{12}|^2}}
{\left(\tilde{C}_{11}+\tilde{C}_{22}+2\sqrt{\tilde{C}_{11}\tilde{C}_{22}-
|\tilde{C}_{12}|^2}\right)^{1/2}}, \label{F_ii_hermit}
\end{equation}
or, in matrix form,
\begin{equation}\label{F_matix_hermit}
\tilde{\mathbf{F}}=
\frac{\tilde{\mathbf{C}}+\sqrt{\det\tilde{\mathbf{C}}}\;\mathbb{I}}{\sqrt{\mbox{tr}
\,\tilde{\mathbf{C}}+2\sqrt{\det\tilde{\mathbf{C}}}}}.
\end{equation}
Below, in section~\ref{Ex-Gen}, we use solutions (\ref{Solution_Ch}) and
(\ref{F_ii_hermit})
for the numerical generation of two
cross-correlated sequences with a given correlation matrix.

%%%%%%%%%%%%%%%%%%%%%%%%%%%%%%%%%%%%%%%%%%%%%%%%%%%%%%%%
\section{Probability distribution function}\label{CharFunc}
%%%%%%%%%%%%%%%%%%%%%%%%%%%%%%%%%%%%%%%%%%%%%%%%%%%%%%%%%

It is well known that most part of the transport properties of
complex random systems are determined by the Fourier transform of
the correlation function. Nevertheless, on frequent occasions we
have to know the PDF of the underlying random sequence. It is just
for that we study the statistical properties of sequences
constructed through the use of the convolution method.

\textbf{1. Weak short-range correlations.}

The normalized filtering function, $\sum_{r=0}^{1} F(r)^2 =
1$, of the form
\begin{equation}\label{weak cor}
F(r)=(1-\frac{\nu^2}{2})\delta_{r,0} + \nu \delta_{r,1}, \quad \quad
|\nu| \ll 1.
\end{equation}
provides a minimal (asymmetric) model governing all the statistical
properties of the sequence with weak correlations.  Using equations~(\ref{x F
alpha}) and (\ref{C sum F}) one readily gets
\begin{equation}\label{Exam Kr weak}
K(r)=\delta_{r,0} + \nu \delta_{|r|,1}.
\end{equation}
The positive values of $\nu$ correspond to the correlation function
describing the sequence with persistent correlations or, in other
words, superdiffusion.
Persistence means an ``attraction'' between the elements of
the same sign and implies superdiffusion, whereas antipersistence means a
``repulsion'' of the elements of the same sign  and is accompanied by subdiffusion.
To demonstrate this, let us introduce an
important statistical characteristic of a random sequence ---
the coordinate variance $D(r)$ for an imaginary Brownian particle
\begin{equation}\label{CoordVar}
D(r)= \langle [x(n+1) + x(n+2)+...+x(n+r)]^2 \rangle.
\end{equation}
Here $x(n+1)$ stands for the length of the first jump, the sum
$x(n+1) + x(n+2)+...+ x(n+r)$ is the coordinate of particle after
$r$ jumps. The variance can be found either by straightforward
calculation (it is simple in this case only) or by ``integration''
of the discrete equation connecting the variance to the correlation
function~\cite{uyakm},
\begin{equation}\label{CorD}
K(r)=\frac{1}{2}[D(r+1) -2 D(r)+D(r-1)], \,\, r\geq 1.
\end{equation}
To proceed, it makes sense to introduce the integrated correlation
function $I(r)$, the first integral of equation (\ref{CorD}), which
satisfies the recurrence relation
\begin{equation}\label{ICF}
I(r+1)=I(r)+K(r), \,r\geq 0.
\end{equation}
The second integral of equation (\ref{CorD}) is
\begin{equation}\label{DoublICF}
D(r+1)=D(r) + 2 I(r+1).
\end{equation}
The last two equations follow from  equation (\ref{CorD}) and
definition (\ref{CoordVar}). Taking into account the equalities
$D(1)=K(0)=1$ (following from equation (\ref{CoordVar}))  and adopting
for convenience of calculations the ``constant of integration'' $D(0)= 0$,
we obtain
\begin{equation}\label{Drweak}
  D(r)=
  \left\{
    \begin{array}{ll}
    r+ 2\nu (r-1) ,& \left| r \right| \geq 1, \\
    0, &\, \,\,r = 0. \\
  \end{array}
    \right.
\end{equation}

We see that the positive values of the parameter $\nu$ yield positive
corrections to the coordinate variance of uncorrelated Brownian motion $D(r)=r$,
i.e., describe
a weak superdiffusion phenomenon, whereas the negative values of
$\nu$ describe a subdiffusion.

Note that the integrated correlation function is suitable in numerical
studies of random processes as a
clear indicator of the correlation length of the sequence; the position
of maximal value of $I(r)$ corresponds to $R_c$.

Now consider the distribution function $\rho_{X}(x)$ of the random
variable $x(n)$ determined by equations~(\ref{weak cor}). When
correlations are short-range and weak, it is not difficult to find the
one-point distribution function of the correlated sequence
$\{ x (n)\}$. Combining equations~(\ref{x F
alpha}) and~(\ref{weak cor}) we get
\begin{equation}\label{x(n)}
x(n) =(1-\nu^2/2)\xi (n) + \nu \xi (n-1).
\end{equation}
Using the well-known formula
\begin{equation}\label{Sum IRV}
\rho _{Y+Z}(x)=\int\limits_{-\infty }^{\infty }\rho _{Y}(x -z )\rho
_{Z}(z )dz,
\end{equation}
expressing the distribution function of the sum of two independent random
variables $Y+Z$ via the convolution of their individual
distributions,
we arrive at the sought result in terms of the uncorrelated PDF
$\rho_{\Xi}(.)$
\begin{equation}\label{PDFweak}
\rho_{X}(x) = \frac{1}{1-\nu^2/2}
\,\rho_{\Xi}\!\left(\frac{x}{1-\nu^2/2}\right)+\frac{\nu^2}{2(1-\nu^2/2)^3}
\rho_{\Xi}{''}\!\left(\frac{x}{1-\nu^2/2}\right).
\end{equation}

Here, the first term is the distribution function of random variable
$(1-\nu^2/2)\xi (n)$, whereas the second one is a small correction
due to the second term in equation~(\ref{x(n)}). The
PDF for $X$ is slightly narrower and steeper than initial
distribution $\rho_{\Xi}(\xi)$ and contains additional narrow and
small humps near the maximum of $\rho_{\Xi}(\xi)$.

Despite the lack of symmetry in the filtering function ~(\ref{weak
cor}) the correlation
function (\ref{Exam Kr weak}) is even. Then, the question arises: which of statistical
characteristics reflect the asymmetry of filtering function?

The lowest by order among the higher order correlation functions is the third-order one:
\begin{equation}\label{Corr 3}
C_{3}(r_{1},r_{2})= \langle x(n) x(n+r_{1})x(n+r_{1}+r_{2})\rangle.
\end{equation}
Straightforward calculation gives
\begin{equation}\label{Corr 01}
C_{3}(0,1)= \langle x(n)^2
x(n+1) \rangle =\langle \xi^3\rangle \nu(1-\nu^2/2)^2,
\end{equation}
\begin{equation}\label{Corr 11}
C_{3}(1,0)= \langle x(n-1) x(n)^2 \rangle = \langle \xi^3 \rangle \nu^2(1-\nu^2/2).
\end{equation}
If PDF of $\xi$ is an even function,  then $\langle\xi^3\rangle =0$. Hence, to
characterize anisotropy of the sequence, we have to turn to the
next, four-point, correlation function:
\begin{equation}\label{Corr 123}
C_{4}(r_{1},r_{2},r_{3})= \langle x(n)
x(n+r_{1})x(n+r_{1}+r_{2})x(n+r_{1}+r_{2}+r_{3}) \rangle;
\end{equation}
\begin{equation}\label{Corr 001}
C_{4}(0,0,1)= \langle x(n)^3 x(n+1) \rangle = \langle \xi^4 \rangle \nu(1-\nu^2/2)^3,
\end{equation}
\begin{equation}\label{Corr 011}
C_{4}(1,0,0)= \langle x(n-1) x(n)^3 \rangle = \langle \xi^4 \rangle \nu^3(1-\nu^2/2).
\end{equation}
Thus, it is clear that the sequence generated by means of asymmetric
filtering function~(\ref{weak cor}) is anisotropic, $C_{3}(0,1)\neq C_{3}(1,0)$ or
$C_{4}(0,0,1)\neq C_{4}(1,0,0)$. The sequences produced by even
filtering functions are isotropic.

Isotropy properties of the multi-step Markov dichotomous sequences
were earlier studied in~\cite{isotr}.

\textbf{2. Long-range correlations.} Now we are interested in
analyzing the case of \emph{long-range correlations} when the
correlation length $R_c$ is large
\begin{equation}\label{LngRCor-def}
R_c\gg1.
\end{equation}

We will show that, if the filtering function is smooth and a large
number of summands contributes to equation~(\ref{C sum F}), the
distribution function $\rho_{X}(x)$ has the \emph{Gaussian} or
\emph{L\'evy} form. This statement is analogous to the Central Limit
Theorem.

The simplest way to demonstrate this is to calculate the
\emph{characteristic function} $\varphi_{X}(t)$ of the random
variable $x(n)$, which is defined by
\begin{equation}\label{CM-ChF-def}
\varphi_{X}(t)\equiv\langle {\exp[it x(n)]}\rangle =
\int_{-\infty}^{\infty}dx\,\rho_{X}(x)\exp(itx).
\end{equation}

From the second equality in definition (\ref{CM-ChF-def}), it
immediately follows that the probability density
$\rho_{X}(x)$ is nothing but the Fourier transform of $\varphi _{X}(t)$,
\begin{equation}\label{rhoB-phiB}
\rho_{X}(x)=\frac{1}{2\pi}\int_{-\infty }^{\infty }dt\,
\varphi_{X}(t)\exp(-itx).
\end{equation}

We substitute the explicit expression (\ref{x F alpha}) for $x(n)$
into definition~(\ref{CM-ChF-def}) of the  characteristic function
$\varphi_{X}(t)$, present the exponential function of the sum of
arguments as a product of exponential functions and take into
account the statistical independence of random variables $\xi(n)$.
This procedure yields
\begin{eqnarray}
 \varphi_{X}(t) &=& \prod_{n'=-\infty}^{\infty}\langle{\exp\left[it
F(n')\xi(n-n')\right]}\rangle
\nonumber
\\
&=&
\prod_{n=-\infty}^{\infty}\int_{-\infty}^{\infty}d\xi
\,
\rho_{\Xi}(\xi)\exp[itF(n)\xi].
\label{CM-phiB}
\end{eqnarray}

Below we will see that the determining contribution into integral
(\ref{rhoB-phiB}) is made by the small values of variable $t$ (due
to a large number of multipliers $F(n)$). At the same time, the
series expansion with respect to the small parameter $t$ depends on
the analytical property of the probability density $\rho_{\Xi}(\xi)$
or, to be more exact, on the behavior of $\rho_{\Xi}(\xi)$ at $|\xi|
\rightarrow\infty$.

\emph{A. Finite dispersion.} Suppose, that $\rho_{\Xi}(\xi)$ is a
rapidly decreasing function, such that the variance $\sigma_{\xi}^2$
exists. In the vicinity of $t=0$, we get
\begin{eqnarray}
 \varphi_{\Xi}(t) &=& \int_{-\infty}^{\infty}d\xi
\,\rho_{\Xi}(\xi)\exp[itF(n)\xi]
\nonumber
\\
&\simeq&
 \int_{-\infty}^{\infty}d\xi
\,\rho_{\Xi}(\xi)[1-t^2 F^2(n)\xi^2] = 1-\frac{1}{2} \sigma_{\xi}^2
F^2(n) t^{2}.
\label{gen var phi}
\end{eqnarray}
Substituting this result (with $\sigma_{\xi}=1)$ into
equation~(\ref{CM-phiB}), after some algebra (typical for the Laplace
method of integral calculations) we have
\begin{eqnarray}
\varphi_{X}(t)
&\simeq&
\!\!\prod_{n=-\infty}^{\infty}\![1-\frac{1}{2}
F^2(n)t^{2}]=\!\prod_{n=-\infty}^{\infty}\!\exp\{\ln[1-\frac{1}{2}
F^2(n)t^{2}]\}
\nonumber
\\
&\!\simeq&
\exp\{-\frac{1}{2}\sum_{n=-\infty}^{\infty}
F^2(n)t^{2}\}.
\label{Fin-Gauss-varphi}
\end{eqnarray}
The characteristic function of this form gives rise to the Gaussian
distribution function~\cite{IKMU,RewUAMM}:
\begin{equation}\label{pho-Gauss}
\rho_{X}(x)=\frac{1}{\sqrt{2\pi\sigma^{2}}}
\exp\left[-\frac{x^{2}}{2\sigma^{2}}\right], \quad
\sigma^{2}=\sum_{n=-\infty}^{\infty}
 F^2(n).
\end{equation}
If $F(n)=1$ for $n=1,2,...,N$ and $F(n)=0$ for $n \leq 0$ and $n > N$, we
recover the well-known result
of the Central Limit Theorem.

We can easily generalize result (\ref{pho-Gauss}) to the sequences $\{x_{j} (n)\}$
generated by equation (\ref{def_x_i}) as a sum of convolutions of
delta-correlated sequences $\{\xi_{j} (n)\}$ with
filtering functions $F_{ij}(n)$. As a consequence of this
calculations for the random variable $x_{i} (n)$,
determined by equation (\ref{def_x_i}), we have the Gaussian distribution function
\begin{equation}\label{VarBunch}
\rho_{X_i}(x_i)=\frac{1}{\sqrt{2\pi\sigma_i^{2}}}
\exp\left[-\frac{x_i^{2}}{2\sigma_i^{2}}\right]
\end{equation}
with the variance
\begin{equation}\label{SigmaBunch}
\sigma^{2}_{i}=\sum_{j=1}^{N}\sum_{n=-\infty}^{\infty}
 F_{ij}^2(n).
\end{equation}

\emph{B. Infinite dispersion.} In the last decades a new class of
systems that do not obey the law of large numbers has emerged [2,3].
The behavior of these systems is dominated by large and rare
fluctuations that are characterized by broad distributions with
power-law tails. The hallmark of these statistical distributions,
commonly referred to as L\'evy statistics [4], is the divergence of
their second and/or first moment.

Suppose $\rho_{\Xi}(\xi)$ to be a slowly decreasing function, such
that property (\ref{118}) does not hold anymore. Then, the large
values of $\xi$ determine the characteristic function
$\varphi_{X}(t)$ for small values of $t$. This type of statements is
known as the Abel - Tauberian theorem for the Fourier transform. Let
us demonstrate this by considering the special form of
$\rho_{\Xi}(\xi)$ - the Student distribution function - generalized
to the fractional value of index $\alpha $,
\begin{equation}\label{17}
\rho_{\Xi}(\xi)=\frac{\Gamma \left( \frac{\alpha +1}{2}\right) }{\sqrt{\pi }%
\Gamma \left( \frac{\alpha }{2}\right) }\frac{b^{\alpha }}{\left(
b^{2}+\xi^{2}\right) ^{\frac{\alpha +1}{2}}},\quad \alpha >0, \quad
b >0.
\end{equation}
Here $\Gamma(.)$ is the gamma function. The characteristic function of
$\rho_{\Xi}(\xi)$ reads
\begin{equation}\label{20}
\varphi _{\Xi}(t)=2\int\limits_{0}^{\infty }\cos \left( t\xi\right)
\rho_{\Xi}(\xi)d\xi=\frac{2}{\Gamma \left( \frac{\alpha }{2}\right)
} \left( \frac{bt}{2}\right)^{\frac{\alpha }{2}} K_{\frac{\alpha
}{2}}\left( bt\right),
\end{equation}
where $K_{a}(.)$ is the modified Bessel function of order $a$.
Taking into account the asymptotic relations for the modified Bessel
function  we obtain in the limit $t\rightarrow 0$
\begin{equation}\label{24}
\varphi _{\Xi}(t)=1\!+\!\frac{\pi }{\Gamma \!\left( \frac{\alpha
}{2}\right) \sin
\pi \alpha /2}\!\left[ \frac{1}{\Gamma \!\left( -\frac{\alpha }{2}+2\right) }%
\left( \frac{bt}{2}\right) ^{2} \!\!\!-\frac{1}{\Gamma \! \left( \frac{\alpha }{2}%
+1\right) }\left( \frac{bt}{2}\right) ^{\alpha }\right]\!.
\end{equation}
We see that $\alpha=2$ is a critical value dividing the asymptotic
behavior of the characteristic function at small values of $t$ into
two regions.

If $\alpha >2$ we can neglect the second term in the square brackets
of equation~(\ref{24}) and, using the recurrence relation for the gamma
function, $\Gamma \left( z+1\right) =z\Gamma \left( z\right) $, we
recover the above obtained result~(\ref{gen var phi}), $
 \varphi_{\Xi}(t)=1- \sigma^2 t^{2}/2, \,\, \sigma
^{2}=b^{2}/(\alpha -2)$. Note that the characteristic function
contains the $\alpha$-independent term $t^{2}$.

Now we are especially interested in the values $\alpha <2$. In this
case we can neglect the first term in the square brackets, so that
we have for the characteristic function
\begin{equation} \label{26}
\varphi _{\Xi_{1}}(t)=1-\frac{\Gamma \left( 1-\frac{\alpha }{2}\right) }{%
\Gamma \left( 1+\frac{\alpha }{2}\right) }\left( \frac{bt}{2}\right)
^{\alpha }, \quad \alpha <2.
\end{equation}
In contrast to the region $\alpha >2$, the exponent of the second
term in the characteristic function is now  $\alpha$-dependent.

Let us consider a more general family of distributions
$\rho_{\Xi}(\xi)$ than the Student one. Assume that
$\rho_{\Xi}(\xi)$ is an even slowly decreasing function with the
asymptotic property
\begin{equation}\label{GenRoXi}
\rho_{\Xi}(\xi) \rightarrow C |\xi|^{-(1+\alpha))} \,\,
%\texttt{at}
\mbox{at}
\,\,|\xi| \rightarrow \infty.
\end{equation}
We can normalize the distribution $\rho_{\Xi}(\xi)$, so that for
small $t$ the  characteristic function has the form
\begin{equation} \label{GenRoXi_t}
\varphi _{\Xi_{1}}(t)=1-t^{\alpha }.
\end{equation}
As an example of such kind of distributions we can take
equation~(\ref{17})\ if we choose the parameters $b$ and $\alpha$
satisfying the equality $2\Gamma \left( 1-\alpha /2 \right)
=b^{\alpha}\Gamma \left( 1+\alpha/2 \right)$. In line with
equations~(\ref{gen var phi}) and~(\ref{Fin-Gauss-varphi}) we obtain the
following results:
\begin{eqnarray} \label{Fin-Levy-varphi}
\varphi_{X}(t)=
\prod_{n=-\infty}^{\infty}\int_{-\infty}^{\infty}d\xi
\,\rho_{\Xi}(\xi)\exp[itF(n)\xi] \!\simeq
\exp\{-\sum_{n=-\infty}^{\infty} F^{\alpha}(n)t^{\alpha}\},
\end{eqnarray}
\begin{equation}\label{28}
\rho_{X}(x)=\frac{1}{\pi }\int\limits_{0}^{\infty }\exp (-\gamma
t^{\alpha })\cos (tx)dt,\quad \gamma =\sum_{n=-\infty}^{\infty}
F^{\alpha}(n).
\end{equation}
Note, all the Gaussian functions ``do not remember'' the form of its
initial distribution $\rho_{\Xi}(\xi)$, whereas the L$\acute{e}$vy
distributions decrease at long distances in the same manner as the
initial ones.

Result (\ref{28}) for the infinite PDF variance sequences
generated by equation (\ref{def_x_i})
is transformed into
\begin{equation}\label{VarBunchLev}
\gamma_{i}=\sum_{j=1}^{N}\gamma_{ij} \,=\,
\sum_{j=1}^{N}\sum_{n=-\infty}^{\infty}
 F^{\alpha}_{ij}(n).
\end{equation}
%

%%%%%%%%%%%%%%%%%%%%%%%%%%%%%%%%%%%%%%%%%%%%%%%%%%%%%%%%%%%%%%%%%%%%%%%
\section{Example of generation.\label{Ex-Gen}}
%%%%%%%%%%%%%%%%%%%%%%%%%%%%%%%%%%%%%%%%%%%%%%%%%%%%%%%%%%%%%%%%%%%%%%%

From the viewpoint of physical applications it is interesting
to consider two delta-correlated sequences with a given
cross-correlation function. For example, let us generate sequences
with correlations given by the following matrix

\begin{equation}\label{example_Cn}
 \mathbf{C}(n)=
\pmatrix{
    \delta_{n,0} \quad & \quad  \frac{1}{2}(\delta_{n,-1}+\delta_{n,1}) \cr
    \frac{1}{2}(\delta_{n,-1}+\delta_{n,1}) \quad & \quad  \delta_{n,0}
    },
    \end{equation}
or, in term of the Fourier transform,
\begin{equation}\label{example_Ck}
\tilde {\mathbf{C}}(k)=
\pmatrix{
    1      \quad & \quad  \cos k \cr
    \cos k \quad & \quad   1
}.
\end{equation}
The Cholesky-like  decomposition of (\ref{example_Ck})
yields
\begin{equation}\label{example_Fk}
\tilde {\mathbf{F}}(k)=
\pmatrix{
        1                \quad & \quad    0\cr
    \cos k    \quad & \quad   i \sin k
}.
\end{equation}
Applying the inverse Fourier transform (\ref{FourTr1}), we can
recover the Cholesky filtering functions $F_{ij}^C(n)$ in real space:
\begin{equation}\label{example_Fn_Ch}
\tilde {\mathbf{F}}^C(n)=
\pmatrix{
\delta_{n,0} \quad & \quad  0\cr
\frac{1}{2}
(\delta_{n,-1}+\delta_{n,1})
 \quad & \quad  \frac{1}{2}(\delta_{n,-1}-\delta_{n,1})
}.
    \end{equation}
Now we can construct numerical sequences $\{x_{1,2}^C(n)\}$ according
to equations (\ref{def_x_i}). In our simple example these sequences
are:
\begin{eqnarray}
x_{1}^C(n) & = & \xi_{1}(n),
\label{ex_x1_Ch}\\
x_{2}^C(n) & = &
\frac{1}{2}[\xi_{1}(n+1)+\xi_{1}(n-1)+\xi_{2}(n+1)-\xi_{2}(n-1)].
\label{ex_x2_Ch}
\end{eqnarray}
Substituting (\ref{ex_x1_Ch}) and (\ref{ex_x2_Ch}) into (\ref{def_ncorr}),
one can see that the correlation properties of the generated
sequences are described by given matrix (\ref{example_Cn}).

Bringing into play solution (\ref{F_ii_hermit}), we construct a new
pair of sequences $\{x_{1,2}^H(n)\}$ correlated in the same way. The
Fourier transforms of the new Hermitian filtering functions are
\begin{eqnarray}
\tilde{F}_{11}^H(k)=
\tilde{F}_{22}^H(k)&=&
\cos{\left( \frac{\pi}{4}-\frac{k}{2} \right)},\\
\tilde{F}_{12}^H(k)=
\tilde{F}_{21}^H(k)&=&
\sin{\left( \frac{\pi}{4}-\frac{|k|}{2} \right)}. \label{F_ij_k_exherm}
\end{eqnarray}
Here the filtering matrix $\mathbf{F}^H$ is Hermitian, therefore its real
entries should be even.
The corresponding filtering functions $F_{ij}^H(r)$ have the form
\begin{eqnarray}
\tilde{F}_{11}^H(n)=
\tilde{F}_{22}^H(n)&= &
    \left\{
        \begin{array}{l}
        A(n) \quad\mbox{for even} \quad n,\\
        0, \quad \mbox{otherwise},
  \end{array}
    \right.
    \\
\tilde{F}_{12}^H(n)=
\tilde{F}_{21}^H(n)&=&
    \left\{
        \begin{array}{l}
        -A(n) \quad\mbox{for odd} \quad n,\\
        0, \quad \mbox{otherwise},
  \end{array}
    \right.\\
\label{F_ij_n_exherm}
\end{eqnarray}
where $A(n)=2\sqrt{2}/{\pi(1-4n^2)}$, and we can generate new
numerical cross-correlated sequences $\{x_{1,2}^H(n)\}$ in
accordance with equations (\ref{def_x_i}). Mention that different
decompositions of the correlation matrix provide the filtering
matrix elements with essentially different analytical properties.

Often the controlling parameters of processes in random systems are
determined  by the Fourier transform of correlation functions of
disorder. By this reason in the plot we present the Fourier
transform of the given cross-correlation function $\tilde C_{12}(k)$
(see matrix (\ref{example_Ck})) and the results of its numerical
calculations with the use of equations (\ref{FourTr1}),
(\ref{def_ncorr}) and (\ref{def_x_i}) for the both pairs of
cross-correlated sequences $\{x_{1,2}^C(n)\}$ and
$\{x_{1,2}^H(n)\}$. The length of the delta-correlated sequences
$\{\xi_{j} (n)\}$ is $10^6$.

\begin{figure}[!h]
\begin{centering}\scalebox{0.7}[0.8]
{\includegraphics{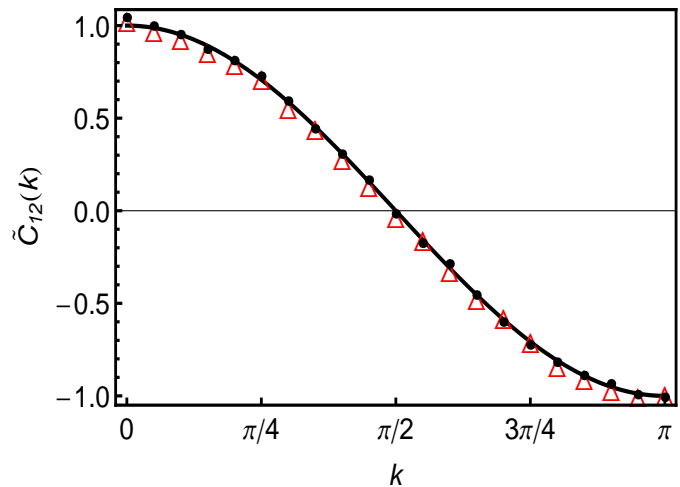}} \caption{Fourier transform of given
cross-correlation function $C_{12}$: solid curve is $\tilde
C_{12}(k)$ entry of matrix (\ref{example_Ck}); points correspond to
numerically calculated cross-correlator of generated sequences
$\{x_{1,2}^C(n)\}$ and triangles stand for numerically calculated
cross-correlator of generated sequences $\{x_{1,2}^H(n)\}$. }
\end{centering}
\end{figure}
%

%%%%%%%%%%%%%%%%%%%%%%%%%%%%%%%%%%%%%%%%%%%%%%%%%%%%%%%%%%%%%%%%%%
\section{Conclusion}\label{Concl}
%%%%%%%%%%%%%%%%%%%%%%%%%%%%%%%%%%%%%%%%%%%%%%%%%%%%%%%%%%%%%%%%%%
In conclusion, let us summarize briefly the main results of the
paper. Despite section 2 is introductory, it contains a few new
results. We clarify the meaning of the filtering function $F(r)$ and
show that it is the value of the cross-correlation function which
describes correlations between the initial white noise and
constructed correlated sequences. This function is determined up to
the gauge factor containing an arbitrary odd function. There is no
restriction on the parity of the filtering function.

In section 3  we present the matrix generalization of the method for
a bunch of $N$ sequences. To construct $N$ cross-correlated
sequences we start with $N$ independent uncorrelated white-noise
random sequences $\{\xi_i(n)\}$. Similarly to the 1-sequence
convolution method we built $N$ cross-correlated sequences
$\{x_{i}(n)\}$ as a sum of convolutions of delta-correlated
sequences $\{\xi_{j}(n)\}$ with filtering functions $F_{ij}(n)$. The
set of these functions is obtained via the factorization of the
Hermitian matrix $\tilde{\mathbf{C}}(k)$ into a product of the
Fourier transforms of the generating function
$\tilde{\mathbf{F}}(k)$ and its Hermitian transpose
$\tilde{\mathbf{F}}^{\dagger}(k)$. Different decompositions of the
correlation matrix are considered: spectral, Cholesky, LDL and
Hermitian ansatz. Explicit expressions for some particular cases are
presented. It was noticed that different decompositions of the
correlation matrix provide the filtering matrix elements with
essentially different analytical properties.

Statistical properties of the sequences constructed by the convolution
method were examined. One-point probability distribution functions
in the cases of weak and strong correlations were studied.
The correlation function, integrated correlation function and the second
integral of the correlation function (the variance of the sum of $L$
random variables) were found for asymmetric weak short-range
correlations in the 1-sequence case. It was shown that the even part of the
filtering function is responsible for generation of isotropic
sequences. It will be interesting to study this phenomenon in the
long-range correlation limit.

If the filtering function is smooth and a large number of summands
contribute to equation~(\ref{C sum F}), the distribution function
$\rho_{X}(x)$ has the Gaussian or L\'evy form.

An example of numerical construction of two correlated chains with a
given correlation matrix was presented. Two different decompositions
of the correlation matrix  were used. It was shown that both of them
give identical numerically reconstructed correlation functions (in
spite of the difference of their analytical properties).

\ack We would like to thank Vekslerchik V E for useful and
stimulating discussions.

%

%%%%%%%%%%%%%%%%%%%%%%%%%%%%%%%%%%%%%%%%%%%%%%%%%%%%%%%%%%%%%%%%%%%%%%%
\begin{center}\textbf{APPENDIX}\end{center}
%%%%%%%%%%%%%%%%%%%%%%%%%%%%%%%%%%%%%%%%%%%%%%%%%%%%%%%%%%%%%%%%%%%%%%%

Here we answer the question how we can generate Fourier harmonics
$\tilde{\xi}(k)$ for random uncorrelated sequence $\{ \xi(n)\}$ of
finite length $N>>1$.
Consider the complex form of the discrete Fourier transform  for $\xi(n)$:
\begin{equation}\label{FourTrComp}
\xi(n) =  \sum_{m=-N+1}^{N-1} \tilde{\xi}(k)\exp (ikn), \,\,
k=k_{m}=\frac{2\pi}{N} m.
\end{equation}
The Fourier coefficients are:
\begin{equation}\label{FourTrCorresp}
\Re \tilde{\xi}(k) = \frac{1}{N}\sum_{n=0}^{N} \xi(n) \cos kn,
\,\,\Im \tilde{\xi}(k) = \frac{1}{N}\sum_{n=0}^{N} \xi(n) \sin kn.
\end{equation}
Here symbols $\Re$ and $\Im$ stand for the real and the imaginary
parts of a complex number. From equations~(\ref{pho-Gauss}),
(\ref{FourTrCorresp}) and formulas $\sum_{n=0}^{N-1} \sin^2 k
n=(1-\delta_{k,0})N/2$,\,\,\,$\sum_{n=0}^{N-1} \cos^2 k n =
(1+\delta_{k,0})N/2$, it follows that the random variables  $\Re
\tilde{\xi}(k)$ and $\Im \tilde{\xi}(k)$ are Gaussian distributed
ones with variances $\sigma^2_{\Re
\tilde{\xi}(k)}=(1+\delta_{k,0})/2N, \,\,\sigma^2_{\Im
\tilde{\xi}(k)}=(1-\delta_{k,0})/2N$.

The values of $\Re \tilde{\xi}(k)$ and $\Im \tilde{\xi}(k)$ for
negative $k$ (after generating $\Re \tilde{\xi}(k)$ and $\Im
\tilde{\xi}(k)$ for $k>0$) have to be determined from relationships:
\begin{equation}\label{FTR}
\Re \tilde{\xi}(-k)=\Re \tilde{\xi}(k),\,\,\,\Im
\tilde{\xi}(-k)=-\Im \tilde{\xi}(k).
\end{equation}

So, instead of generating a sequence $\{ \xi(n)\} $ of uncorrelated
random numbers and calculating then their Fourier transform
coefficients, we can generate directly complex random numbers
$\tilde{\xi}(k)=\Re \tilde{\xi}(k) + i\Im \tilde{\xi}(k)$.

To formulate the inverse statement let us consider the discrete Fourier
transform for $\xi(n)$:
\begin{equation}\label{FourTrReal}
\xi(n)=a_0+ \sum_{m=1}^{N-1}( a_k\cos kn + b_k\sin kn), \,\,
k=k_{m}=\frac{2\pi}{N}m,
\end{equation}
and suppose that the Fourier components $a_k$ and $b_k$ are
independent and identically distributed variables with the variances
$\sigma^2_{a_k}=\sigma^2_{b_k}=1/N$. We conclude that the random
variables $\xi(n)$ are Gaussian distributed ones with equal
variances $\sigma^2_{\xi(n)}=1$. This follows immediately from
equation~(\ref{pho-Gauss}).
%
%%%%%%%%%%%%%%%%%%%%%%%%%%%%%%%%%%%%%%%%%%%%%%%%%%%%%%%%%%%%%%%%%%%
\section*{References}
%%%%%%%%%%%%%%%%%%%%%%%%%%%%%%%%%%%%%%%%%%%%%%%%%%%%%%%%%%%%%%%%%%%

\end{document}